\documentclass[11pt]{article}
\usepackage[margin=1in]{geometry}
\usepackage{amsmath,amssymb,amsthm,booktabs,xspace,xcolor}
\usepackage[hidelinks]{hyperref}

\newtheorem{lemma}{Lemma}
\newtheorem{theorem}{Theorem}

\newcommand{\ropebwt}{\texttt{ropebwt3}\xspace}
\newcommand{\K}{\mathcal{K}}

\newcommand{\gb}{\raisebox{\dimexpr\aboverulesep+\belowrulesep+\cmidrulewidth}[0pt][0pt]{(GB)}}

\title{Adaptive BML}
\author{Travis Gagie\\[1ex]
\small Center for Biotechnology and Bioengineering (CeBiB), Chile}
\date{}

\begin{document}
\maketitle

\begin{abstract}
\noindent
Maximal exact matches (MEMs) between a pattern and a text are widely used in bioinformatics, and for tasks such as
classifying reads it can be enough to know the few longest MEMs of each read.  Boyer--Moore--Li (BML) finds all
the MEMs of length at least a given threshold $L$ while ignoring much of the pattern, and $k$-mer based breaking
(KeBaB) lets BML ignore even more, but both need $L$ to be chosen in advance.  We describe adaptive BML (ABML),
which is given a parameter $\tau$ instead, raises $L$ as it runs and returns $\tau$ longest MEMs.  ABML combines
naturally with KeBaB and the early-stopping rule proposed with it, which had not been implemented.  ABML is simple:
our implementation is a small patch to Li's \ropebwt.  On about 580,000 Oxford Nanopore reads and an index of about 5,900 bacterial genomes, finding 5 longest MEMs
of each read with KeBaB and ABML took 11 minutes on one thread of an old desktop, against 1.9 hours for
finding all the MEMs of length at least 40 with BML and 8.0 hours with forward--backward.
\end{abstract}

\section{Introduction}

A maximal exact match (MEM) of a pattern $P [1..m]$ with respect to a text $T [1..n]$ is a substring $P [i..j]$
that occurs in $T$ and cannot be extended to the left or to the right in $P$ and still occur in $T$ (\ropebwt calls these super-maximal exact matches).  Long MEMs are
informative and short MEMs are mostly noise, so applications usually keep only the MEMs of length at least some
threshold $L$.  Li's forward--backward algorithm~\cite{Li2012} finds all the MEMs with an FM-index~\cite{FM2005} and
then the short MEMs are discarded.  Boyer--Moore--Li (BML)~\cite{Gagie2024BML} is given $L$ and finds exactly the
MEMs of length at least $L$, usually without reading all of $P$; Li implemented it in \ropebwt~\cite{Li2024}, which now uses it by default.  With $k$-mer based breaking (KeBaB)~\cite{BrownEtAl2025}, we first use a Bloom
filter~\cite{Bloom1970} to cut $P$ into substrings called pseudo-MEMs that together contain all the MEMs of length
at least $L$, and then search only the pseudo-MEMs.  Using $k$-mers to avoid examining most of the input when only long MEMs are wanted is older than KeBaB: to find MEMs between two genomes, copMEM~\cite{GrabowskiBieniecki2019} samples $k$-mers from both genomes with coprime steps, and copMEM2~\cite{GrabowskiBieniecki2023} also filters the sampled $k$-mers with a Bloom filter.

For some tasks we do not need all the MEMs of length at least $L$.  Brown et al.~\cite{BrownEtAl2025} observed that
metagenomic classification can often be done from only a few long MEMs of each read.  They pointed out that, if we
search the pseudo-MEMs in non-increasing order by length, then we can stop when we have found $t$ MEMs longer than
the next pseudo-MEM~\cite[Proposition~2]{BrownEtAl2025}.  (With the lengths of the pseudo-MEMs as upper bounds, this is the stopping rule of Fagin et al.'s threshold algorithm for top-$k$ queries~\cite{FaginEtAl2001}.)  That early-stopping rule needs a modified MEM-finder,
however, so they left implementing it as future work: in one experiment they simulated it, and in another they
searched only the $t$ longest pseudo-MEMs.  Searching only the $t$ longest pseudo-MEMs is fast but it is a heuristic: a long pseudo-MEM need not contain a long MEM, so the MEMs found
are not necessarily the longest MEMs of the read.

In this note we show how to find the longest MEMs exactly and quickly.  Given $\tau$, we want $\tau$ MEMs of $P$
such that no other MEM of $P$ is longer than any of them (or all the MEMs of $P$, if there are fewer than $\tau$).
Our main observation is that BML stays correct if $L$ increases while it runs, so once we have found $\tau$ MEMs we
can keep $L$ at 1 plus the length of the $\tau$th longest MEM found so far.  We call the result \emph{adaptive BML
(ABML)} and describe it in Section~\ref{sec:abml}.  The idea of raising the threshold during the search goes back to Goga et al.~\cite{GogaEtAl2024}, who gave an
LCP-based algorithm for finding all the MEMs of at least a given length and observed that changing the threshold
during the query lets that algorithm find all the longest common substrings of $P$ and $T$, which they tested.  The
appendix of the preprint of the BML paper~\cite{Gagie2024arXiv}, which is not in the published
version~\cite{Gagie2024BML}, did the same with BML: there we kept $L$ at 1 more than the length of the longest MEM
found so far and counted the backward steps on synthetic data.  Here we generalize from the longest MEMs to $\tau$
longest MEMs for any $\tau$, prove ABML correct, implement it in a widely used MEM-finder and test it on real
data.  Raising a threshold to the score of the $k$th best answer found so far, in order to skip candidates that cannot beat it, is standard in top-$k$ document retrieval~\cite{TurtleFlood1995,BroderEtAl2003}.  In Section~\ref{sec:kebab} we combine ABML with KeBaB: the
early-stopping rule becomes a comparison of the next pseudo-MEM's length with the current value of $L$, the value
of $L$ carries over from one pseudo-MEM to the next, and if the pseudo-MEMs run out before we have found $\tau$
MEMs then we fall back on searching all of $P$.
In Section~\ref{sec:experiments} we report an experiment with a patched \ropebwt, in which finding 5 longest MEMs
with KeBaB and ABML was 42 times faster than forward--backward, 10 times faster than BML and 4.9 times faster than
KeBaB with BML.

\section{Preliminaries}
\label{sec:preliminaries}

We assume $T$ is indexed with an FM-index, so that for any substring $P [i..j]$ we can find the longest suffix of
$P [i..j]$ that occurs in $T$ by matching backward from $P [j]$, and the longest prefix of $P [i..j]$ that occurs
in $T$ by matching forward from $P [i]$.  (In practice an index of $T$ and its reverse complement supports both
directions.)  To simplify the presentation we assume every character of $P$ occurs in $T$; otherwise we can cut $P$
at the characters that do not occur in $T$.

\paragraph{Forward--backward.}
A MEM starts at $P [1]$.  If a MEM starts at $P [s]$ then we find it by finding the longest prefix $P [s..e]$ of
$P [s..m]$ that occurs in $T$.  If $e = m$ then we are done.  Otherwise, the next MEM starts at $P [s']$, where
$P [s'..e + 1]$ is the longest suffix of $P [s..e + 1]$ that occurs in $T$, and we continue from $s'$.  We call
finding $e$ and then $s'$ a \emph{round} of forward--backward.

\paragraph{BML.}
Suppose we want only the MEMs of length at least $L$, and that we know no MEM of length at least $L$ that we have
not reported starts before $P [s]$; initially $s = 1$.  If fewer than $L$ characters remain, that is, $s + L - 1 >
m$, then we are done.  Otherwise, we find the longest suffix $P [s'..s + L - 1]$ of the window $P [s..s + L - 1]$
that occurs in $T$.  If $s' > s$ then we continue from $s'$.  If $s' = s$ then a MEM of length at least $L$ starts
at $P [s]$, so we run a round of forward--backward from $s$, report the MEM that the round finds and continue from
the start of the next MEM.

\paragraph{KeBaB.}
Choose $k \leq L$ and let $\K$ be a superset of the $k$-mers in $T$; in practice $\K$ is the set of $k$-mers that a
Bloom filter built from $T$ accepts.  The \emph{pseudo-MEMs} of $P$ are the maximal substrings of $P$ that contain
only $k$-mers in $\K$ and have length at least $L$.  Brown et al.\ stated, without proof and for $L > k$, that the pseudo-MEMs contain all the MEMs of $P$ of length at
least $L$~\cite[Proposition~1]{BrownEtAl2025}.  For searching the pseudo-MEMs to be correct we also need the
converse, that a long enough MEM of a pseudo-MEM is a MEM of $P$, so we state and prove both directions; $k \leq L$
is enough.

\begin{lemma}
\label{lem:kebab}
For any $L' \geq L$, the MEMs of length at least $L'$ of the pseudo-MEMs with respect to $T$ are the same as the
MEMs of length at least $L'$ of $P$ with respect to $T$.
\end{lemma}

\begin{proof}
Suppose $P [i..j]$ is a MEM of $P$ of length at least $L'$.  All the $k$-mers in $P [i..j]$ occur in $T$ and
$P [i..j]$ has length at least $L$, so $P [i..j]$ is contained in some pseudo-MEM $P [h..\ell]$.  Extending
$P [i..j]$ in either direction in $P [h..\ell]$ gives a substring of $P$ that does not occur in $T$, so $P [i..j]$
is a MEM of $P [h..\ell]$.

Now suppose $P [i..j]$ is a MEM of some pseudo-MEM $P [h..\ell]$, of length at least $L' \geq k$.  If $i > h$ then
$P [i - 1..j]$ does not occur in $T$.  If $i = h = 1$ then $P [i..j]$ cannot be extended to the left in $P$.  If
$i = h > 1$ then, since $P [h..\ell]$ is maximal, the $k$-mer $P [h - 1..h + k - 2]$ is not in $\K$ and so does not
occur in $T$; since $j \geq h + k - 1$, that $k$-mer is contained in $P [i - 1..j]$, which therefore does not occur
in $T$.  Symmetrically, $P [i..j]$ cannot be extended to the right in $P$ and still occur in $T$, so $P [i..j]$ is
a MEM of $P$.
\end{proof}

\section{Adaptive BML}
\label{sec:abml}

ABML is BML with two changes: it records the MEMs it finds instead of reporting them at once and, each time it
records a MEM, if it has recorded at least $\tau$ MEMs then it sets $L$ to 1 plus the length of the $\tau$th
longest MEM recorded so far.  Figure~\ref{fig:abml} shows pseudo-code.  The value of $L$ never decreases.  When ABML
finishes, it returns the $\tau$ longest MEMs it has recorded, or all of them if it has recorded fewer than $\tau$.

\begin{figure}[t]
\centering
\begin{minipage}{.9\textwidth}
\hrule\smallskip
\textbf{ABML}$(P [1..m], \tau, L)$
\begin{enumerate}\itemsep 0pt
\item $s \leftarrow 1$
\item while $s + L - 1 \leq m$:
\begin{enumerate}\itemsep 0pt
\item find the longest suffix $P [s'..s + L - 1]$ of $P [s..s + L - 1]$ that occurs in $T$
\item if $s' > s$ then $s \leftarrow s'$ and continue
\item find the longest prefix $P [s..e]$ of $P [s..m]$ that occurs in $T$, and record the MEM $P [s..e]$
\item if at least $\tau$ MEMs are recorded then $L \leftarrow 1 + {}$the length of the $\tau$th longest of them
\item if $e = m$ then stop
\item find the longest suffix $P [s'..e + 1]$ of $P [s..e + 1]$ that occurs in $T$, and $s \leftarrow s'$
\end{enumerate}
\item return the $\tau$ longest MEMs recorded (all of them, if there are fewer than $\tau$)
\end{enumerate}
\smallskip\hrule
\end{minipage}
\caption{Adaptive BML.  Without step~(d) and with the MEMs reported in step~(c), it is BML.}
\label{fig:abml}
\end{figure}

\begin{lemma}
\label{lem:invariant}
When ABML finishes, every MEM of $P$ that it has not recorded is shorter than the final value of $L$.
\end{lemma}

\begin{proof}
Since $L$ never decreases, it is enough to show that every MEM that ABML passes without recording it is shorter
than the value of $L$ at that time.  ABML moves $s$ to the right in three ways.

First, suppose ABML moves from $s$ to $s' > s$ in step~(b), and a MEM of length at least $L$ starts at $P [i]$ with
$s \leq i < s'$.  Since that MEM has length at least $L$ and $i \geq s$, it ends at or after $P [s + L - 1]$, so it contains $P [i..s + L - 1]$, which is then a suffix of the
window that occurs in $T$ and is longer than $P [s'..s + L - 1]$, a contradiction.

Second, suppose ABML moves from $s$ to $s'$ in step~(f).  It has recorded the MEM that starts at $P [s]$.  If a
MEM starts at $P [i]$ with $s < i < s'$ then, since one MEM cannot contain another, it ends after $P [e]$ and so
contains $P [i..e + 1]$, which is then a suffix of $P [s..e + 1]$ that occurs in $T$ and is longer than
$P [s'..e + 1]$, a contradiction.  Similarly, if ABML stops in step~(e) then no MEM starts after $P [s]$.

Third, when the loop ends with $s + L - 1 > m$, any MEM that starts at or after $P [s]$ has length at most
$m - s + 1 < L$.
\end{proof}

\begin{theorem}
\label{thm:abml}
Suppose ABML starts with $L = L_0$.  If $P$ has fewer than $\tau$ MEMs of length at least $L_0$ then ABML returns
all of them.  Otherwise, ABML returns $\tau$ MEMs of $P$ such that no other MEM of $P$ is longer than any of them.
\end{theorem}

\begin{proof}
Every MEM that ABML records has length at least the value of $L$ at that time, which is at least $L_0$.  If ABML
records fewer than $\tau$ MEMs then $L$ stays $L_0$ and, by Lemma~\ref{lem:invariant}, ABML records every MEM of
length at least $L_0$.  Otherwise, the final value of $L$ is $1 + \lambda$, where $\lambda$ is the length of the
$\tau$th longest MEM recorded, and by Lemma~\ref{lem:invariant} every MEM not recorded has length at most
$\lambda$.
\end{proof}

Starting with $L_0 = 1$ gives $\tau$ longest MEMs of $P$ whatever their lengths.  If several MEMs are tied for
$\tau$th longest then ABML returns only some of them; if we want all the MEMs tied for $\tau$th longest as well,
then in step~(d) we set $L$ to the length of the $\tau$th longest MEM recorded so far instead of 1 plus that
length, and return every recorded MEM at least that long.  ABML starts more slowly than BML with a well-chosen
$L$, because its windows are short until it has found $\tau$ MEMs, but it needs no knowledge of how long the MEMs
will be, and when $P$ has many long MEMs it ends with a larger $L$ than we would dare to choose in advance.

\section{KeBaB with adaptive BML}
\label{sec:kebab}

Suppose we choose $k \leq L_0$, find the pseudo-MEMs of $P$ for $k$ and $L_0$, and sort them into non-increasing
order by length, as for Brown et al.'s early-stopping rule~\cite[Proposition~2]{BrownEtAl2025}.  We search the
pseudo-MEMs in that order with ABML, starting with $L = L_0$ and carrying the
recorded MEMs and the value of $L$ from one pseudo-MEM to the next.  Then one of three things happens:
\begin{enumerate}
\item the next pseudo-MEM is shorter than the current value of $L$: we stop and return the $\tau$ longest MEMs
recorded;
\item the pseudo-MEMs run out and we have recorded at least $\tau$ MEMs: we return the $\tau$ longest of them;
\item the pseudo-MEMs run out and we have recorded fewer than $\tau$ MEMs: we discard them and run ABML on all of
$P$ starting with $L = 1$.
\end{enumerate}

\begin{theorem}
\label{thm:kebab}
The procedure above returns $\tau$ MEMs of $P$ such that no other MEM of $P$ is longer than any of them, or all the
MEMs of $P$ if there are fewer than $\tau$.
\end{theorem}

\begin{proof}
The value of $L$ is always at least $L_0 \geq k$, so by Lemma~\ref{lem:kebab} every MEM recorded while searching a
pseudo-MEM is a MEM of $P$, and every MEM of $P$ of length at least $L_0$ is a MEM of exactly one pseudo-MEM, which
is at least as long as that MEM.  Let $L^*$ be the value of $L$ when we stop searching pseudo-MEMs.  By
Lemma~\ref{lem:invariant}, applied to each pseudo-MEM we searched, every MEM of $P$ of length at least $L_0$ that
lies in a pseudo-MEM we searched and is not recorded is shorter than $L^*$.  In case~1 the pseudo-MEMs we did not
search are all shorter than $L^*$, and so are the MEMs they contain.  In cases~1 and~2, therefore, every MEM of $P$
not recorded is shorter than $L^*$ or shorter than $L_0$.  In case~1 we have $L^* > L_0$, since every pseudo-MEM has
length at least $L_0$, so we have recorded at least $\tau$ MEMs; hence in cases~1 and~2 the $\tau$th longest MEM
recorded has length $\lambda \geq L_0$ and, if $L^* > L_0$, then $L^* = 1 + \lambda$.  Either way no MEM not recorded is
longer than $\lambda$.  In case~3 the claim follows from Theorem~\ref{thm:abml}.
\end{proof}

Case~3 occurs exactly when $P$ has fewer than $\tau$ MEMs of length at least $L_0$, so some of the MEMs we want are
shorter than $L_0$ and KeBaB cannot help to find them.  If we want only MEMs of length at least $L_0$ then we skip
the fallback and return what we have recorded.  The condition $k \leq L_0$ matters: a MEM of a pseudo-MEM that is
shorter than $k$ need not be a MEM of $P$.  Brown et al.'s rule stops when $t$ MEMs longer than the next pseudo-MEM
have been found; since we do not need all the MEMs tied for $\tau$th longest, case~1 also stops when $\tau$ MEMs
at least as long as the next pseudo-MEM have been found.

\section{Experiment}
\label{sec:experiments}

We implemented ABML and the procedure of Section~\ref{sec:kebab}, including early stopping, as a patch to \ropebwt
(commit b408339, four commits after release 3.10-r281) that adds options to its \texttt{mem} command: one gives $\tau$, and another gives a file with
each read's pseudo-MEMs, as pairs of positions in the read, in non-increasing order by length.  A small patch to
KeBaB writes that file.  The patch keeps the $\tau$ largest lengths found so far for the
current read in a min-heap, so resetting $L$ takes $O (\log \tau)$ time per MEM.  The patch, its tests and our scripts are available at
\url{https://github.com/TravisGagie/Adaptive_Boyer-Moore-Li}.

We used the setting of Brown et al.'s experiment with \ropebwt~\cite{BrownEtAl2025}.  The text is the collection
that Ahmed et al.~\cite{AhmedEtAl2023} built from all the strains in RefSeq of the 7 bacterial species in the Zymo
mock community, which Brown et al.\ report as 5,867 genomes; with plasmids it has 14,280 sequences and 27~GB in
all.  The patterns are the 581,802 Oxford Nanopore reads that Brown et al.\
used as positive reads, with average length 19,692~bp by our count and 11.5~GB in all; they are from SRA run SRR11071395
(experiment SRX7711546) of the Zymo High Molecular Weight Mock Microbial Community, sequenced by Kovaka et
al.~\cite{KovakaEtAl2021}.  We ran KeBaB with $k = 20$ and $L_0 = 40$, values that Brown et al.\ used with these reads; we did not tune them.  Since KeBaB needs $L_0 \geq k$, it cannot start from $L = 1$, so we start from $L = 1$ only with ABML alone.  Everything ran
on one thread of a 7-year-old desktop with an Intel Core i7-8700 processor (6 cores, 3.2~GHz) and 16~GB of RAM, under Ubuntu 26.04.  The data structures of KeBaB and of
\ropebwt were never in memory at the same time, and the times for KeBaB with BML or ABML are the sums of the times
of the two steps, including the time for KeBaB to scan the reads.  KeBaB with BML uses more memory than the other rows
only because unmodified \ropebwt reads its queries in batches of 100 million bases, and a batch of pseudo-MEMs has many
more records than a batch of reads; for KeBaB with ABML, our patch reads whole reads and keeps the pseudo-MEMs as pairs of positions.  The first three rows are for unmodified \ropebwt and the last three are for \ropebwt with our patch.  Every time is a median, of ten runs for KeBaB with ABML and of three runs for the other rows, except that the time for forward--backward is from a single run.  With the option for ABML off, the patched \ropebwt gives the same output as unmodified \ropebwt in times within 0.3\% (6,963 seconds for BML and 3,340 seconds for KeBaB with BML), but with KeBaB its peak memory is 2\% higher, so we report the unmodified tool for those rows.  We
checked that every run with ABML returned,
for every read, MEMs with exactly the lengths of the $\tau$ longest MEMs found by searching the whole read with
$L = 1$.

\begin{table}[t]
\centering
\caption{Finding MEMs between 581,802 Oxford Nanopore reads and a collection reported as 5,867 bacterial genomes with \ropebwt: total time in
seconds, the size of the index on disk (with KeBaB, of the index and the Bloom filter) and the peak memory (with KeBaB, of
\ropebwt's step, which is the larger), with 1~GB being $10^9$ bytes.  ABML
from $L = 40$ falls back on searching the whole read from $L = 1$ when it finds fewer than $\tau$ MEMs of length at
least 40, with or without KeBaB.}
\label{tab:results}
\medskip
\begin{tabular}{lrrrrcc}
 & \multicolumn{4}{c}{total time (s)} & disk & RAM \\
\cmidrule(lr){2-5}
all MEMs of length at least 40 & & & & & \gb & \gb \\
\midrule
forward--backward       & \multicolumn{4}{c}{28,909} & 0.77 & 1.07 \\
BML                     & \multicolumn{4}{c}{\phantom{2}6,943} & 0.77 & 1.07 \\
KeBaB + BML             & \multicolumn{4}{c}{\phantom{2}3,330} & 1.04 & 1.67 \\
\midrule
$\tau$ longest MEMs & $\tau = 1$ & $\tau = 5$ & $\tau = 10$ & $\tau = 20$ & & \\
\midrule
ABML from $L = 1$              & 1,685 & 3,014 & 4,011 & 5,493 & 0.77 & 0.99 \\
ABML from $L = 40$             & 1,645 & 2,759 & 3,499 & 4,469 & 0.77 & 0.99 \\
KeBaB + ABML from $L = 40$     &   383 &   684 &   957 & 1,419 & 1.04 & 0.99
\end{tabular}
\end{table}

Table~\ref{tab:results} shows the results.  The first three rows find the same 87,996,223 MEMs of length at least
40.  For $\tau = 5$, ABML alone from $L = 1$ is 2.3 times faster than BML with $L = 40$, even though it is given no
threshold, and KeBaB with ABML is 4.4 times faster than ABML alone and 42 times faster than forward--backward.
The advantage shrinks as $\tau$ grows, as we would expect, but for $\tau = 20$ KeBaB with ABML is still 4.9 times
faster than BML.

ABML does well here because most reads have many long MEMs, so $L$ rises quickly: the median length of a read's
longest MEM is 197 and the median length of its 10th longest MEM is 115.  The fallback is rarely needed: 6,585
reads (1.13\%) have no MEM of length at least 40, 12,385 (2.13\%) have fewer than 5 and 16,518 (2.84\%) have fewer
than 10.  Those reads are searched twice when we start from $L = 40$, which costs less than it saves on the other
reads.

This experiment has limits.  It shows that finding $\tau$ longest MEMs is fast, not that $\tau$ longest MEMs are enough for classification: for that we rely on Brown et al.'s observation, and we have not measured accuracy.  When several MEMs are tied for $\tau$th longest, which of them are returned depends on the order of the search, so the output is determined only up to such ties.  We used one dataset, with long reads that mostly have many long MEMs; with short reads, or reads with few long MEMs, $L$ rises more slowly and ABML saves less.

We expect larger speedups from running the reads, or the pseudo-MEMs, in parallel, since the index is static.  ABML
and early stopping apply to any MEM-finder that implements BML, with any index that supports matching in both
directions.

\section*{Acknowledgments}

Many thanks to my coauthors on the KeBaB paper~\cite{BrownEtAl2025}, in which the idea of using only a few long
MEMs of each read arose, and to Heng Li for \ropebwt.

\paragraph{Use of AI.}
The patch to \ropebwt and its tests were written by Anthropic's Claude, which also ran the experiments under my
direction, found errors in earlier versions of my descriptions and proofs, made suggestions about the
implementation, and helped to draft this note.  I have checked the note and take responsibility for its contents.

\end{document}